\documentclass[aps,prl,reprint,superscriptaddress,nofootinbib]{revtex4-2}

\usepackage{amsmath,amssymb,amsthm}

\usepackage{graphicx}
\usepackage{hyperref}
\hypersetup{hidelinks}

\newcommand{\ep}[1]{\mathrm{e}^{#1}}
\newcommand{\iu}{\mathrm{i}}

\newcommand{\bbR}{\mathbb{R}}

\newcommand{\bbZ}{\mathbb{Z}}

\newcommand{\caA}{\mathcal{A}}

\newcommand{\str}{^\ast}

\newcommand{\ind}{\mathrm{ind}}

\newtheoremstyle{prltheorem}
{0.5\baselineskip}{0.5\baselineskip}{\normalfont}{}% spaces and body font
{\itshape}{}{0.5em}% head font, punctuation, and spacing
{\thmname{#1}\thmnumber{ #2}\thmnote{ (#3)}---}
\theoremstyle{prltheorem}
\newtheorem{theorem}{Theorem}

\usepackage{tikz}
\usetikzlibrary{patterns,arrows,calc}

\definecolor{edgecol} {RGB}{ 43, 90,140}  % dark blue: all rules and arrows
\definecolor{fluxcol} {RGB}{  0,  0,  0}  % arrows stay black
\definecolor{hatchcol}{RGB}{110,150,190}
\definecolor{disclite}{RGB}{228,239,250}  % annulus, far side
\definecolor{discdark}{RGB}{196,217,238}  % annulus, near side
\definecolor{surfhi}  {RGB}{240,247,253}  % cylinder highlight
\definecolor{surfmid} {RGB}{203,224,242}  % cylinder, far edge
\definecolor{surfdark}{RGB}{160,191,223}  % cylinder, near edge
\definecolor{boredark}{RGB}{146,180,214}  % inside of the tube, top
\definecolor{borelite}{RGB}{194,216,237}  % inside of the tube, bottom
\definecolor{surfcol} {RGB}{213,230,245}  % flat fallback fill
\definecolor{holecol} {RGB}{186,210,235}

\definecolor{cellfill}{RGB}{240,246,252}   % empty orbital
\definecolor{occfill} {RGB}{214,231,247}   % singly occupied orbital
\definecolor{occfill2}{RGB}{186,213,240}   % doubly occupied orbital
\definecolor{ballcol} {RGB}{200, 40, 40}   % particles

\tikzset{
  rim/.style      ={draw=edgecol,line width=1.1pt},
  rib/.style      ={draw=edgecol!45,line width=0.4pt},
  fluxline/.style ={draw=fluxcol,line width=1.1pt},
  pump/.style     ={draw=black,line width=1.0pt,->,>=latex},
  maparrow/.style ={draw=black,line width=1.0pt,->,>=latex},
  cell/.style      ={draw=edgecol,line width=0.5pt,fill=cellfill},
  cellocc/.style   ={draw=edgecol,line width=0.5pt,fill=occfill},
  cellocctwo/.style={draw=edgecol,line width=0.5pt,fill=occfill2},
  mono/.style      ={draw=edgecol,line width=0.9pt,fill=occfill,
                     rounded corners=1pt},
  sep/.style       ={draw=edgecol,line width=0.5pt}, % as the void boxes
  ball/.style      ={circle,draw=ballcol!70!black,fill=ballcol,
                     inner sep=0pt,minimum size=4.0pt},
  shiftarrow/.style ={->,>=latex,line width=0.6pt,draw=black},
  oparrow/.style    ={->,>=latex,line width=0.7pt,draw=black},
  pastearrow/.style ={<-,>=latex,line width=0.7pt,draw=black},
  seam/.style       ={draw=edgecol!70,line width=0.5pt,densely dashed},
  ruler/.style      ={font=\tiny\sffamily,text=black!65},
  lab/.style        ={font=\normalsize,text=black},
  note/.style       ={font=\small,text=black},
}

\newcounter{sitec}
\newcommand{\occrow}[1]{%
  \setcounter{sitec}{0}%
  \edef\occlist{#1}%
  \foreach \n in \occlist {%
    \pgfmathtruncatemacro{\sx}{\number\value{sitec}}%
    \ifnum\n=0\relax \draw[cell] (\sx,0) rectangle (\sx+1,1);\fi
    \ifnum\n=1\relax \draw[cellocc] (\sx,0) rectangle (\sx+1,1);\fi
    \ifnum\n=2\relax \draw[cellocctwo] (\sx,0) rectangle (\sx+1,1);\fi
    \ifnum\n=1\relax \node[ball] at (\sx+0.5,0.5) {};\fi
    \ifnum\n=2\relax \node[ball] at (\sx+0.5,0.29) {};
                     \node[ball] at (\sx+0.5,0.71) {};\fi
    \stepcounter{sitec}%
  }%
}

\newcounter{monoj}
\newcommand{\tilingrow}[3][0]{%
  \setcounter{monoj}{0}%
  \edef\bseq{#3}%
  \foreach \bv in \bseq {\stepcounter{monoj}\xdef\blast{\bv}}%
  \pgfmathtruncatemacro{\NN}{\number\value{monoj}}%
  \ifnum#1>0\relax \pgfmathtruncatemacro{\LL}{#1}%
  \else            \pgfmathtruncatemacro{\LL}{\NN*#2+\blast}\fi
  \pgfmathtruncatemacro{\Lm}{\LL-1}%
  \foreach \s in {0,...,\Lm}{\draw[cell] (\s,0) rectangle (\s+1,1);}%
  \setcounter{monoj}{0}%
  \foreach \bv in \bseq {%
    \stepcounter{monoj}%
    \pgfmathtruncatemacro{\jj}{\number\value{monoj}}%
    \pgfmathtruncatemacro{\st}{(\jj-1)*#2+\bv}%
    \draw[mono] (\st,0) rectangle (\st+#2,1);
    \foreach \d in {1,...,\the\numexpr#2-1\relax}{%
      \draw[sep] (\st+\d,0) -- (\st+\d,1);}%
    \node[ball] at (\st+0.5,0.5) {};
  }%
}

\newcommand{\orbruler}[1]{%
  \pgfmathtruncatemacro{\Lm}{#1-1}%
  \foreach \s in {0,...,\Lm}{%
    \node[ruler,anchor=north] at (\s+0.5,-0.12) {\s};}%
}

\begin{document}

\title{Fractional Hall Conductance of Laughlin States from a Topological Index}

\author{Sven Bachmann}
\email{sbach@math.ubc.ca}
\affiliation{Department of Mathematics, The University of British Columbia,
Vancouver, British Columbia V6T 1Z2, Canada}

\author{Severin Schraven}
\email{severin.schraven@tum.de}
\affiliation{Department of Mathematics, Technical University of Munich,
85748 Garching, Germany}

\author{Jacob Shapiro}
\email{jacobshapiro@princeton.edu}
\affiliation{Department of Mathematics, Princeton University,
Princeton, New Jersey 08544, USA}

\author{Simone Warzel}
\email{simone.warzel@tum.de}
\affiliation{Department of Mathematics, Technical University of Munich,
85748 Garching, Germany}
\affiliation{Department of Physics, Technical University of Munich,
85748 Garching, Germany}
\affiliation{MCQST, 
80799 M\"unchen, Germany}
\date{\today}

\begin{abstract}
    We provide a rigorous Laughlin-pump argument that establishes the fractional quantization of the Hall conductance directly from the Laughlin state as a topological index.  Our proof uses a recently defined index of a pair of pure states together with an infinite matrix product state analysis that is rigorous on a sufficiently thin cylinder. By virtue of the connection to an index, our result implies, in particular, that Laughlin states are representatives of topologically stable fractional quantum Hall phases.
\end{abstract}

\maketitle

\begin{figure}[ht]
\centering
\resizebox{\columnwidth}{!}{%
\begin{tikzpicture}[x=0.45cm,y=0.45cm]

% =====================================================================
%  (a) Cylinder: domain of w = exp(gamma z)
%       local frame: x' along the axis, y' across it
% =====================================================================
  \def\Lc{7.0}    % length of the tube
  \def\Rc{1.15}   % radius (semi-axis across the tube)
  \def\ec{0.42}   % foreshortening of the end caps
  \def\tilt{32}   % tilt of the tube axis, in degrees

  \begin{scope}[shift={(-4.8,-1.6)},rotate=\tilt]

    % ---- shaded cylindrical surface ---------------------------------
    \shade[bottom color=surfdark,
           middle color=surfhi,
           top color=surfmid]
       (0,\Rc) -- (\Lc,\Rc)
       arc [start angle=90, end angle=-90,
            x radius=\ec, y radius=\Rc]
       -- (0,-\Rc)
       arc [start angle=270, end angle=90,
            x radius=\ec, y radius=\Rc]
       -- cycle;

    % ---- circumferential ribs ---------------------------------------
    \foreach \f in {0.18,0.36,0.54,0.72,0.90}{%
      \draw[rib] ({\f*\Lc},\Rc)
        arc [start angle=90, end angle=-90,
             x radius=\ec, y radius=\Rc];}

    % ---- silhouette lines -------------------------------------------
    \draw[rim] (0,\Rc) -- (\Lc,\Rc);
    \draw[rim] (0,-\Rc) -- (\Lc,-\Rc);

    % ---- far opening -------------------------------------------------
    \draw[rim] (\Lc,\Rc)
       arc [start angle=90, end angle=-90,
            x radius=\ec, y radius=\Rc];

    % ---- near opening ------------------------------------------------
    \shade[top color=boredark, bottom color=borelite]
          (0,0) ellipse [x radius=\ec, y radius=\Rc];
    \draw[rim] (0,0) ellipse [x radius=\ec, y radius=\Rc];

    % ---- flux line entering the near end -----------------------------
    \draw[fluxline] (-2.1,0) -- (\ec,0);

    % ---- pumped charge along the cylinder ----------------------------
    \draw[pump]
      ({0.20*\Lc},{-0.34*\Rc})
      --
      ({0.56*\Lc},{-0.34*\Rc});

    \node[lab,anchor=south,inner sep=3pt]
      at ({0.30*\Lc},{-0.10*\Rc})
      {$\Delta Q$};

    % ---- flux line leaving the far end -------------------------------
    \draw[fluxline] (\Lc+\ec,0) -- (\Lc+2.1,0);
    \draw[fluxline,->,>=latex]
      (\Lc+1.3,0) -- (\Lc+2.1,0);

    \coordinate (fluxlab)
      at (\Lc+1.05,0);

  \end{scope}

% Upright flux label outside the rotated cylinder scope
\node[lab,anchor=south east,inner sep=2pt]
  at ($(fluxlab)+(-0.02,0.28)$)
  {$\phi$};

% Conformal-map label
\node[lab,anchor=south]
  at (5.15,2.05)
  {$z\longmapsto \mathrm{e}^{\gamma z}$};

% ---- cylinder radius ------------------------------------------------
%\draw[<->,>=latex,line width=0.55pt]
%  (0,0) -- (0,\Rc)
%  node[midway,anchor=west,inner sep=2pt]
 % {$\gamma^{-1}$};
  
% =====================================================================
%  Conformal map: cylinder -> annulus
% =====================================================================
  \draw[maparrow]
    (3.5,1.15)
    to[out=32,in=152]
    (6.4,1.35);

%  \node[lab,anchor=south]
 %   at (4.95,2.05)
  %  {$z\longmapsto w=e^{\gamma z}$};

% =====================================================================
%  (b) Annulus: image under w = exp(gamma z)
% =====================================================================
  \begin{scope}[shift={(12.7,0)}]

    \def\Rx{6.0}   \def\Ry{2.55}   % outer ellipse semi-axes
    \def\rx{1.05}  \def\ry{0.45}   % inner ellipse semi-axes
    \def\ttop{3.7} \def\tbot{-3.1} % extent of the flux line

    % ---- annular surface ---------------------------------------------
    \shade[top color=disclite,
           bottom color=discdark,
           even odd rule]
      (0,0) ellipse [x radius=\Rx, y radius=\Ry]
      (0,0) ellipse [x radius=\rx, y radius=\ry];

    % ---- images of the cylinder ribs --------------------------------
    \foreach \k in {1,...,5}{%
      \pgfmathsetmacro{\ax}{\rx+\k*(\Rx-\rx)/6}%
      \pgfmathsetmacro{\ay}{\ry+\k*(\Ry-\ry)/6}%
      \draw[rib]
        (0,0) ellipse [x radius=\ax, y radius=\ay];
    }

    % ---- inner and outer rims ----------------------------------------
    \draw[rim]
      (0,0) ellipse [x radius=\Rx, y radius=\Ry];
    \draw[rim]
      (0,0) ellipse [x radius=\rx, y radius=\ry];

    % ---- flux line through the hole ---------------------------------
    \draw[fluxline]
      (0,\tbot) -- (0,-\Ry);
    \draw[fluxline]
      (0,-\ry) -- (0,\ttop);
    \draw[fluxline,->,>=latex]
      (0,\ttop-0.9) -- (0,\ttop);

    \node[lab,anchor=west,inner sep=3pt]
      at (0.18,\ttop-0.45)
      {$\phi$};

    % ---- radially pumped charge -------------------------------------
    \def\tq{22}
    \coordinate (qa)
      at ({\rx*cos(\tq)},{\ry*sin(\tq)});
    \coordinate (qb)
      at ({\Rx*cos(\tq)},{\Ry*sin(\tq)});

    \draw[pump] (qa) -- (qb);

    \node[lab,anchor=north,inner sep=2pt]
      at ($(qa)!0.72!(qb)+(0.25,-0.20)$)
      {$\Delta Q$};

  \end{scope}

\end{tikzpicture}%
}
\caption{The conformal equivalence between a finite cylinder and a
planar annulus under $z\longmapsto\mathrm{e}^{\gamma z}$. The
cylinder radius is $R_{\rm cyl}=\ell_B/\gamma$; extending the cylinder
infinitely along its axis gives the punctured plane.}
\label{fig:conformal}
\end{figure}
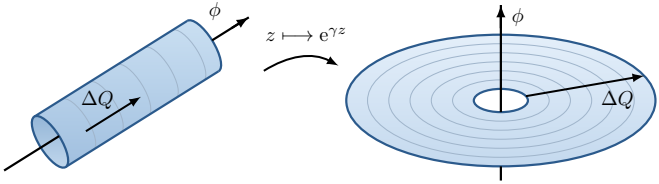

The Hall conductance has a particularly direct interpretation in Laughlin's
flux-insertion thought experiment~\cite{Laughlin1981}. In a planar, Corbino geometry,
inserting one flux quantum through the hole pumps charge from one boundary to
the other  (cf.~Fig.~\ref{fig:conformal}), yielding 
\begin{equation}\label{def:HC}
\begin{aligned}
    \frac{h}{e}\,\sigma_{\rm H}=\ &\text{charge transported from the inner}\\
    &\text{boundary to the outer boundary.}
\end{aligned}
\end{equation}
The history of theoretical arguments surrounding the fractional quantization of $ \sigma_{\rm H} $ is rich. 
Following Laughlin's proposal \cite{Laughlin1983FQHE} of an incompressible state at filling $1/q$ with fractional excitations, Halperin \cite{Halperin1983Theory} argued that the corresponding Hall response is $\sigma_{\rm H}=e^2/(qh)$. Laughlin's flux-insertion argument was recast in gauge-invariant and topological terms  \cite{TaoWu1984,NiuThoulessWu1985}, and recent works have revisited this topic by formulating $ \sigma_{\rm H} $  geometrically through the Laughlin bundle over Aharonov--Bohm fluxes~\cite{KlevtsovZvonkine2022}, or by showing how $  \sigma_{\rm H} $ can be extracted directly from a single gapped many-body wavefunction~\cite{FanSahayVishwanath2023}. Laughlin's topological charge pump has been realized experimentally in an atomic Hall cylinder \cite{FabreEtAl2022} and in a thin-film magnetic heterostructure \cite{Kawamura2023}.

In this Letter, we recast the result of Laughlin's original argument as a robust topological invariant associated with a pair of states \cite{Bachmann2026Index} and compute its value for the explicit fermionic Laughlin state on a sufficiently thin cylinder, relating it to the Tao-Thouless construction~\cite{PhysRevB.28.1142}. We prove that $q$ flux insertions carry a many-body index equal to one, concluding with~\eqref{def:HC} that $2\pi\sigma_{\rm H}=1/q$ (now switching to natural units). This follows directly from the structure of the Laughlin state, without assuming a spectral gap or an abstract topologically ordered ground-state manifold~\cite{HastingsMichalakis2015,PhysRevLett.120.096601,Kapustin_2020,BachmannBolsDeRoeckFraas2021}, and identifies the Laughlin states as mere representatives of topological FQH phases. 
%
%%%%%%%%%%%%%
\paragraph{The index of a pair of states.}
For non-interacting electrons, the charge transport by flux-insertion can be expressed as the
index of a pair of Fermi projections~\cite{Bellissard1988,PhysRevLett.65.2185,Avron1994,BellissardVanElstSchulzBaldes1994}.  This index is integer-valued and topologically stable.  A central point of the present discussion is
that a similar structure survives in an interacting many-body setting and that it yields fractional charge transport. The
single-particle projections are replaced by many-body pure states, and the topological  index measures the net charge separating the two states.  We
briefly recall the formalism developed in \cite{Bachmann2026Index}.

We consider a pair $(\omega_1,\omega_2)$ of pure states on the CAR algebra of observables $\caA$. The many-body analogue of two projections having trace-class difference is the following.  We say
that $\omega_1$ and $\omega_2$ belong to the same \emph{selection sector} if
\begin{equation*}
    \omega_2(A)=\omega_1(U\str A U).
\end{equation*}
for some unitary $U\in\caA$. Physically, this means that the two states differ locally, but they asymptotically agree at infinity.

Charge conservation enters through the $U(1)$ gauge action $\rho_\phi$ on $\caA$, with $\rho_{2\pi}=\mathrm{id}$.
One may choose a sequence of increasingly large local charge
operators $(Q_n)$ in $\caA$ such that \cite{C3}
\begin{equation*}
    \frac{\rm d}{\rm d\phi}\rho_\phi(A)
    =
    \lim_{n\to\infty}\iu[Q_n,\rho_\phi(A)].
\end{equation*}
We assume that the states under consideration are gauge invariant, i.e.,
\begin{equation*}
    \omega(\rho_\phi(A))=\omega(A)
\end{equation*}
for every $A\in\caA$ and every $\phi\in\mathbb{R}$.

The charge-transport index associated with a pair of gauge-invariant pure states in the same selection sector is
\begin{equation}\label{eq:ind equal Delta Q}
   \ind(\omega_1,\omega_2)
   =
   \lim_{n\to\infty}\bigl(\omega_1(Q_n)-\omega_2(Q_n)\bigr)\in\mathbb{Z}.
\end{equation}
Thus the index measures the asymptotic charge deficiency of $\omega_2$ relative to $\omega_1$. In this setting,
$\omega_2(A)=\omega_1(U\str A U)$ for some unitary $U\in\caA$, and one equivalently has~\cite{Bachmann2026Index} 
\begin{equation*}
    \ind(\omega_1,\omega_2)
    =\iu\frac{\rm d}{\rm d\phi}\omega_1(U\str\rho_\phi(U))\vert_{\phi = 0}\in\mathbb{Z}.
\end{equation*}
The important facts are that this quantity is \emph{integer-valued}, independent of
the particular choice of the approximating charges $Q_n$ and of the
implementing unitary $U$, and additive $\ind(\omega_1,\omega_3)
    =
    \ind(\omega_1,\omega_2)+\ind(\omega_2,\omega_3)$. Most importantly for the quantum Hall effect, it is robust. On the one hand,
\begin{equation}\label{eq:Automorphism invariance}
    \ind(\omega_1,\omega_2) = \ind(\omega_1\circ\beta,\omega_2\circ\beta)
\end{equation}
for any automorphism $\beta$ commuting with $\rho_\phi$. On the other hand,
\begin{equation}\label{eq:stability}
    \ind(\omega_1,\omega_2)=0
    \qquad\text{whenever}\qquad
    \Vert\omega_1-\omega_2\Vert<2.
\end{equation}
In other words, an integer charge cannot change under a sufficiently small
many-body deformation.  This is the many-body topological input that we shall
combine with Laughlin's flux-insertion argument.

%%%%%%%%%%%%%
\paragraph{Flux piercing in a Hall sample.}

We now connect this abstract index to the familiar physics of flux insertion.
On the plane, insertion of one flux quantum at the origin is represented at
the one-particle level by the Laughlin flux operator, namely multiplication by
\begin{equation*}
    \ep{\iu\mathrm{arg}(z)}
\end{equation*}
on $L^2(\bbR^2)$, see Fig.~\ref{fig:conformal}.
The standard lowest-Landau-level quasihole construction associated with a
flux quantum at the origin instead amounts to multiplying the wavefunction by $z$. We use this familiar correspondence as
motivation for the orbital shift~\cite{C1}. Hence inserting
$k$ flux quanta act on the holomorphic part of the $N$-particle Laughlin wavefunction as
\begin{equation}\label{eq:flux insertion on Laughlin}
\Psi_{N}(z_1,\ldots,z_N)\mapsto
\prod_{\ell=1}^N z_\ell^k
\prod_{1\leq j<k\leq N}(z_j-z_k)^q.
\end{equation}
The physical effect is particularly transparent in the occupation-number
basis: every occupied orbital is shifted by $k$ sites.
 The case $k=q$ is special, yielding precisely $\Psi_{N+1,q}(0,z_1,\ldots,z_N)$ on the RHS of \eqref{eq:flux insertion on Laughlin}.
Thus, inserting $q$ flux quanta produces the $(N+1)$-particle Laughlin state with one additional fluxon 
pinned at the origin.  This observation is the physical origin of the unit
charge that will appear in the many-body index in Theorem~\ref{thm:Hall} below.

For the rigorous analysis we pass to the conformally
equivalent cylinder through $z\mapsto\ep{\gamma z}$, with $\gamma$ the magnetic length divided by the
cylinder radius, taken large
enough for the quantitative estimates of~\cite{schraven2026fractional}. We moreover simultaneously switch to half-infinite states. The one-particle Hilbert space of the LLL is then identified with
$\ell^2(\bbZ)$ and $\caA$ is the corresponding CAR algebra.  We denote by
$\omega_q$ the infinite-volume Laughlin state at maximal filling, which is empty sufficiently far to the left; namely $\omega_q$ emerges as the unique $N\to\infty$ limit of $\Vert \Psi_{N}\Vert^{-2}\langle \Psi_{N}|\cdot|\Psi_{N}\rangle$; $\omega_q$ is pure \cite{schraven26b}.

Let $\tau$ denote the right translation automorphism of $\caA$,
\begin{equation*}
    \tau(c_j)=c_{j-1}\,,
\end{equation*}
where $c_j,c_j\str$ are the annihilation and creation operators of LLL cylinder orbital
$j$.  A $k$-fold flux inserted state is therefore represented simply by
$\omega_q\circ\tau^k$.  The first key result identifies exactly when the
flux-inserted state remains in the same many-body selection sector.

\begin{theorem}\label{thm:selection sectors}
    If $ \gamma $ is large enough, the states $\omega_q$ and $\omega_q\circ\tau^k$ lie in the same
    selection sector if and only if $k$ is an integer multiple of $q$.
\end{theorem}
This theorem already contains the characteristic fractional structure of the
Laughlin state.  Its physical content is easiest to see in the
Tao--Thouless limit of an infinitely thin cylinder~\cite{PhysRevB.28.1142} where the state
approaches the quasifree crystalline occupation pattern
$\cdots 000000|  1\,0\cdots 0\,1\,0\cdots 0\,1\cdots $,
with one occupied orbital every $q$ sites.  A translation by $q$ orbitals
changes only the boundary of the half-infinite configuration and hence is a
local modification; the two states agree at infinity.  By contrast, a translation by
$1,\ldots,q-1$ changes the occupation pattern all the way to infinity.
Since two pure states on a CAR algebra belong to the same selection sector precisely when they agree at infinity
\cite[Cor.~2.6.11]{bratteli1987operator}, the theorem is immediate in
the Tao--Thouless picture.  The iMPS estimates below show that this distinction
survives away from the strict product-state limit, justifying the $\gamma\to\infty$ heuristics.

The second key result is that the pair
$(\omega_q,\omega_q\circ\tau^q)$ in the same selection sector carries precisely one unit of charge:
\begin{theorem}\label{thm:Hall}
    If $\gamma$ is large enough,
\begin{equation}\label{eq:fractionalHC}
    2\pi q\sigma_{\rm H}(\omega_q)=\ind(\omega_q,\omega_q\circ\tau^q)=1.
    \end{equation}
\end{theorem}
The importance of this identity is that $\sigma_{\rm H}(\omega_q)$ is a many-body topological invariant: It is integer-valued and perturbatively stable. The asymptotic charge deficiency after $q$ flux insertions is one unit of charge. We remark that the threshold for $\gamma$ is explicit, and that it is necessary to justify the approximating of $\omega_q$ by the Tao-Thouless state where explicit calculations are tractable.

The proof of Theorems~\ref{thm:selection sectors} and \ref{thm:Hall} will be
given after the iMPS estimates have been introduced.  Conceptually, however,
the mechanism is already visible: the Laughlin state has a $q$-orbital
periodicity at infinity, $q$ flux insertions return the state to the same
selection sector, and the corresponding many-body index detects one unit of
transported charge.

Theorem~\ref{thm:Hall} is complementary to the general quantization theorems in~\cite{HastingsMichalakis2015,Bachmann2018,Kapustin_2020,BachmannBolsDeRoeckFraas2021,bachmann2025tensor}. There, quantization follows from the assumption of a spectral gap, complemented by a local topological order condition, or the assumption of invertibility, or the finiteness of the anyonic tensor category. In contrast, our result follows directly from the structure of the Laughlin wavefunction in the thin-cylinder regime. In particular, it does not require the a priori assumption of a gap. 

%%%%%%%%%%%%%
\paragraph{Topological invariance.}

If $\beta$ is any automorphism that commutes with both $\rho_\phi$ and flux insertion $\tau$, then $(\omega_q\circ\beta,\omega_q\circ\beta\circ\tau^q)$ are in the same selection sector and
\begin{equation*}
\sigma_{\rm H}(\omega_q\circ\beta) = \sigma_{\rm H}(\omega_q)
\end{equation*}
by~(\ref{eq:Automorphism invariance}). Denoting $\alpha = \beta\circ\tau^q\circ\beta^{-1}\circ\tau^{-q}$, the same conclusion holds if $(\omega,\omega\circ\alpha)$ are in the same selection sector, namely 
$(\omega_q\circ\alpha)(A) = \omega_q(W\str A W)$, 
and if $W$ is gauge-invariant, $\rho_\phi(W) = W$. This is in particular the case if $\beta(A) = V\str A V$, namely it is local, with $W = V\tau^q(V\str)$. In general, $W$ is only such that $\rho_\phi(W) = \ep{\iu n\phi} W$ for some integer $n$, in which case
\begin{equation*}
\sigma_{\rm H}(\omega_q\circ\beta) = \sigma_{\rm H}(\omega_q) + \frac{n}{2\pi q}.
\end{equation*}

We may also consider a continuous deformation $\beta_s$ that commutes with $\rho_\phi$, namely $s\mapsto\beta_s(A)$ is norm continuous for any observable $A$, and define
$  \omega_q(s)=\omega_q\circ\beta_s $.
Whenever the deformed pair $(\omega_q(s),\omega_q(s)\circ\tau^q)$ remains in the same selection sector, the stability~\eqref{eq:stability} of the many-body index implies
\begin{equation*}
    \sigma_{\rm H}(\omega_q(s))
    =
    \sigma_{\rm H}(\omega_q(0)),
\end{equation*}
provided 
$
    \Vert \omega_q\circ\alpha_t - \omega_q\circ\alpha_s\Vert <2 $ 
whenever $|s-t|$ is small.

Thus the value $1/q$ is not a peculiarity of an exactly solvable wavefunction: it is tied to an integer-valued many-body topological invariant and therefore persists under charge-conserving deformations that do not change the asymptotic sector.

\begin{figure}[ht]
\centering
\resizebox{\columnwidth}{!}{%
\begin{tikzpicture}[x=0.50cm,y=0.50cm]

  \def\NS{16}      % orbitals drawn: 0,...,15
  \def\rowa{0}     % baseline of row 1
  \def\rowb{-3.1}  % baseline of row 2
  \def\rowc{-6.2}  % baseline of row 3
  \def\kshift{3}   % the shift, in orbitals

  % ---------------- row 1: the root partition ------------------------
  \begin{scope}[shift={(0,\rowa)}]
    \tilingrow[\NS]{3}{0,0,0,0,0}
    \node[lab,anchor=east] at (-0.5,0.5) {$\lambda^{(3)}_5$};
    \node[note,anchor=west] at (\NS+0.6,0.5) {$(0,3,6,9,12)$};
  \end{scope}

  % ---------------- row 2: a squeezed partition ----------------------
  \begin{scope}[shift={(0,\rowb)}]
    \occrow{0,1,0,1,0,1,0,0,0,0,1,1,0,0,0,0}
    \node[lab,anchor=east] at (-0.5,0.5) {$\mu$};
    \node[note,anchor=west] at (\NS+0.6,0.5) {$(1,3,5,10,11)$};
  \end{scope}

  % ---------------- row 3: the shifted partition ---------------------
  \begin{scope}[shift={(0,\rowc)}]
    \occrow{0,0,0,0,1,0,1,0,1,0,0,0,0,1,1,0}
    \orbruler{\NS}
    \node[lab,anchor=east] at (-0.5,0.5) {$T^{3}\mu$};
    \node[note,anchor=west] at (\NS+0.6,0.5) {$(4,6,8,13,14)$};
  \end{scope}

  % ---------------- squeezing, row 1 -> row 2 ------------------------
  \draw[oparrow] (3.2,\rowa-0.35) -- (3.2,\rowb+1.35);
  \node[note,anchor=west] at (3.5,{(\rowa+\rowb+1)/2}) {squeezing};

  % ---------------- the shift, row 2 -> row 3 ------------------------
  \foreach \s in {1,3,5,10,11}{%
    \draw[shiftarrow] ({\s+0.5},\rowb-0.35)
                   -- ({\s+0.5+\kshift},\rowc+1.35);}
  \node[note,anchor=west] at (\NS+0.6,{(\rowb+\rowc+1)/2})
       {shift by $3$};

\end{tikzpicture}%
}
\caption{In the occupation picture, with red balls symbolizing particles, two elementary squeezings map the root with $q=3$ and $b=0, N = 5 $ on the top line to the second line. The third line is a shift of the previous one, which corresponds to flux insertion. }
\label{fig:squeezing} 
\end{figure}
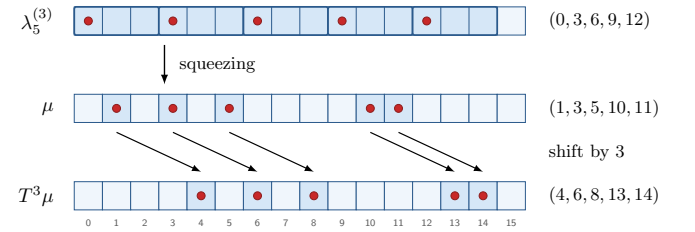

\paragraph{iMPS representation and its main properties.}
For sufficiently large $\gamma$, the Laughlin state and its lower density
descendants admit a controlled \cite{schraven2026fractional} infinite matrix product state (iMPS)
representation~\cite{estienne2013fractional,ZaletelMongPollmann2013,EstienneRegnaultBernevig2015}. This polymer representation associates
with each root partition
\[
    \lambda_b^{(q)}(N)
    =
    \bigl(b_1,b_2+q,\ldots,b_N+q(N-1)\bigr)
\]
the holomorphic part of a Laughlin-type wavefunction
\begin{equation*}
\begin{aligned}
    &\Psi_{b,N}(z_1,\dots,z_N)
    \propto
    m_b(e^{\gamma z_1},\dots,e^{\gamma z_N})
    \\
    &\mkern170mu\times
    \prod_{1\leq i<j\leq N}
    \left(e^{\gamma z_i}-e^{\gamma z_j}\right)^q 
\end{aligned}
\end{equation*} where $q\in\mathbb{N}$ specifies the basic density, $b=(b_1,\cdots,b_N)$ is a tuple of shifts compared to that density, and $ m_b $ is an associated symmetric polynomial. The choice $b=\mathbf{0}_N$ gives the maximally-filled Laughlin state on the cylinder, while general
$b$ introduces holes into its root configuration. The choice $b=k\mathbf{1}_N$ corresponds to a uniform shift of all orbitals by $k$ and will be of interest below. Altogether,
these functions span the zero-mode space of the Haldane
pseudopotential~\cite{Haldane1983} corresponding to a given value of $ q $.
Building on~\cite{di1994laughlin,dunne1993slater,PhysRevLett.94.026802,PhysRevB.86.245305,estienne2013fractional,konig2017matrix,jansen2009symmetry,
nachtergaele2021spectral}, the  new expansion \cite{schraven2026fractional} expresses such functions in terms of the the occupation Slater determinant states $\Phi_\mu$ corresponding to the a partition $\mu$ recording the occupation of one-particle orbitals 
\begin{align*}
    \varphi_k(z)
    =
    \sqrt{\frac{\gamma}{2\pi^{3/2}}}\,
    e^{-\gamma^2 k^2/2}\,
    e^{\gamma z k},
    \qquad k\in\bbZ , 
\end{align*}
and reads
\begin{equation}
\Psi_{b,N}
    = 
    \sum_{\mu\preceq\lambda_b^{(q)}(N)}
    w_b(\mu) \ \Phi_\mu,
    \qquad
    w_b\bigl(\lambda_b^{(q)}(N)\bigr)=1 .
\label{expansion}
\end{equation}
Here 
$\mu\preceq\lambda_b^{(q)}(N)$ means that $\mu$ is obtained from the
root partition by inward squeezings, cf.~Fig.~\ref{fig:squeezing}. Since the expansion coefficient of 
the root partition $\lambda_b^{(q)}$ dominates in the thin cylinder limit, $\Psi_{b,N}$ may be viewed as the root occupation state $\Phi_{\lambda_b^{(q)}(N)}$ dressed by quantum
fluctuations. 

For the arguments below, two quantitative consequences of the iMPS
representation~\eqref{expansion} are decisive.  Writing
$\widehat\Psi_{b,N}:=\Psi_{b,N}/\|\Psi_{b,N}\|$ and $\odot$ the normalized antisymmetrized tensor
product, fixed by
$
\Phi_\mu\odot\Phi_\nu=\Phi_{\mu\cup\nu}$
whenever the occupied orbitals of $\nu$ lie strictly to the right
of those of $\mu$, which in the occupation picture just corresponds to the concatenation of orbitals, the first key result from \cite[Thm~1.7]{schraven2026fractional} is approximate factorization
across any cut:
\begin{equation}\label{eq:cut}
    \left\|
    \widehat\Psi_{b,N_1+N_2}
    -
    \widehat\Psi_{b_1,N_1}
    \odot
    \widehat\Psi_{b_2,N_2}
    \right\|
    \leq
    \frac{2e^{-C_q(\gamma)q/2}}
         {1-e^{-C_q(\gamma)q}} .
\end{equation}
Here the root data are split as
$b=(b^{(1)},b^{(2)})$ at the chosen cut. The estimate is uniform in the
position of the cut and in the positions of the holes. This is a strong statement: in the thin-cylinder regime, where $ C_q(\gamma) = \gamma^2+\mathcal{O}_q(1) $, the
many-body wavefunction is uniformly close to a tensor product across any
chosen cut.  Equivalently, the entanglement generated across the cut is
quantitatively small.

The second input is uniform exponential clustering. For
$   \omega_{b,N}(A)
    :=
    \langle\widehat\Psi_{b,N},
    A \ \widehat\Psi_{b,N}\rangle $, 
one has
\begin{equation}\label{eq:clustering}
\begin{aligned}
&\left|\omega_{b,N}(AB)-\omega_{b,N}(A)\omega_{b,N}(B)\right|
\\
&\;\leq C\|A\|\|B\|
\exp\!\left[-\frac{C_q(\gamma)}{6}
\operatorname{dist}\!\left(\operatorname{supp}A,\operatorname{supp}B\right)
\right].
\end{aligned}
\end{equation}
where $C<\infty$ is independent of $b$ and $N$
\cite[Thm.~1.8]{schraven2026fractional}. Earlier work established
a weak form in a very thin-cylinder regime
\cite{jansen2009symmetry}; the present bounds are quantitative and, crucially
for our application, uniform enough to control flux-inserted states and holes.

Finally, the renewal-theoretic normalization asymptotics
\cite[App.~A]{schraven2026fractional} yield control on the normalization, 
\begin{equation}
    \frac{\|\Psi_{N-1}\|^2}{\|\Psi_N\|^2}
    \longrightarrow r,
    \qquad 0<r<1 .
\label{norms}
\end{equation}
This ratio identifies the local boundary correction used in the proof
of Theorem~\ref{thm:selection sectors}, which we spell out next.

We first treat the basic case $k=q$. 
A $q$-fold translation removes precisely the first occupied root orbital.
Using \eqref{expansion}, we conclude $c_0 \Psi_{\mathbf{0},N} = \Psi_{q \mathbf 1, N-1}$ and hence by \eqref{norms} one obtains for any observable $B\in\caA$
\begin{align}\label{eq:shift-boundary}
  \omega_q(c_0^* B c_0)
  &=\lim_{N\to\infty}
  \frac{\Vert \Psi_{N-1}\Vert^2}{\Vert \Psi_N\Vert^2}
  \frac{\langle \Psi_{N-1},\tau^q(B)\Psi_{N-1}\rangle}
       {\Vert \Psi_{N-1}\Vert^2}
  \notag\\
  &=r\,(\omega_q\circ\tau^q)(B).
\end{align}
The important point is that the effect of the global $q$-fold shift is represented on the state $\omega_q$ by the local boundary operator $\frac{1}{\sqrt{r}}c_0$. In fact, setting $B=\mathbb{I}$, we may extract $r = \omega_q(c_0^*c_0)$. Since $\omega_q$ is even, only the even part of $B$ contributes, and therefore
\[
    \omega_q(c_0^*Bc_0)
    =
    \omega_q(c_0^*c_0 B).
\]
Now specialize to $B$ supported in $[L,\infty)$. The operator $c_0^*c_0$ is localized at the left boundary, whereas $B$ is
supported at distance at least $L$.  Exponential clustering
\eqref{eq:clustering} consequently gives
\[
\begin{aligned}
    \omega_q(c_0^*c_0B)
    &=\omega_q(c_0^*c_0)\omega_q(B)
      +O(\|B\|e^{-cL})\\
    &=r\,\omega_q(B)+O(\|B\|e^{-cL}).
\end{aligned}
\]
Together with \eqref{eq:shift-boundary},
\begin{equation*}
    \left|
    \omega_q\circ\tau^q(B)-\omega_q(B)
    \right|
    \leq
    C\|B\|e^{-cL}.
\end{equation*}
Hence the two states become indistinguishable at infinity.  They therefore
belong to the same selection sector. The same conclusion holds for every $k=nq$, $n\in\bbZ$, by iteration.

It remains to exclude the other congruence classes modulo $q$.  Write
$k=nq+\ell$ with $0<\ell<q$.  Since the $nq$-part does not change the
selection sector, it is enough to consider the case $n=0$.  By the cut estimate~\eqref{eq:cut},  the vector $\Psi_{N}/\Vert \Psi_{N}\Vert$ is well-approximated by $\frac{\Psi_{\ell}}{\Vert \Psi_{\ell}\Vert} \odot \frac{\Psi_{1}}{\Vert \Psi_{1} \Vert} \odot \frac{\Psi_{N-\ell-1}}{\Vert \Psi_{N-\ell-1}\Vert}$, up to errors which are small for large $\gamma$, uniformly in $\ell, N$. A similar factorization holds for the state after $k$ flux insertion, which is built from $\lambda_{k \mathbf{1}}^{(q)}(N)$. We conclude that, on the one hand, $\omega_q(c_{qj}^* c_{qj}) \sim 1$ while $\omega_q\circ\tau^k(c_{qj}^* c_{qj}) \sim 0$ for all $j$, for large enough $\gamma$, i.e.
\begin{equation} \label{exp numb part}
    \omega_q(n_{mq+\ell}) \sim \delta_{\ell, 0}.
\end{equation}
Hence these two states differ at infinity and therefore lie in different selection sectors concluding the proof  of Theorem~\ref{thm:selection sectors}.

Finally, the following short argument yields the quantization of the Hall conductance, Theorem~\ref{thm:Hall}. Since $\omega_q$ is obtained as the infinite-volume limit of states with fixed
particle number, it is gauge-invariant; it is moreover pure \cite{schraven26b}. The index is therefore defined for
the pair in the same selection sector
$(\omega_q,\omega_q\circ\tau^q)$.

Using local charge operators
\[
Q_L=\sum_{i=-L}^{L}n_i,
\]
and
$\tau^q(n_i)=n_{i-q}$, we obtain
\begin{align*}
    \ind&(\omega_q,\omega_q\circ\tau^q)
    =
    \lim_{L\to\infty}
    \Bigl(
        \omega_q(Q_L)
        -
        (\omega_q\circ\tau^q)(Q_L)
    \Bigr)
    \\
    &=
    \lim_{L\to\infty}
    \omega_q\bigg(
        \sum_{i=L-q+1}^{L}n_i
    \bigg)
    -
    \lim_{L\to\infty}
    \omega_q\bigg(
        \sum_{i=-L-q}^{-L-1}n_i
    \bigg)
    \\
    &=1-0=1.
\end{align*}
The last equality uses that the half-infinite Laughlin state is
asymptotically empty to the left and has one particle per $q$
consecutive orbitals to the right by \eqref{exp numb part}. Hence insertion of $q$ flux quanta transports one
unit of charge. By additivity of flux insertion, the charge transported per
single flux quantum is therefore given by~\eqref{eq:fractionalHC}.\\

\paragraph{Discussion.}
We have expressed Laughlin's flux-insertion response through a
 many-body topological index and shown, for sufficiently thin cylinders, that one
flux quantum transports charge $1/q$.  The present argument directly uses the form of the $ q $-Laughlin state and does not rely on an assumed many-body spectral gap of parent Hamiltonians. Our many-body index of a pair of states is simple and also bypasses Green function and linear response approaches~\cite{PhysRevLett.131.236601}.

The stability of the index implies in particular that the Hall conductance remains $1/q$ for any state that is asymptotically equal at infinity to the maximally filled Laughlin state.

Extending the analysis beyond the thin-cylinder large $\gamma$ regime, and to more general fractional quantum Hall states, is a
natural direction for future work.

\medskip
This work was supported by the DFG under grant TRR 352--Project-ID 470903074 (SS, SW) and  EXC-2111 -- 390814868 (SW). SB acknowledges financial support from NSERC of Canada. JS was supported in part by NSF grant DMS-2510207.

\medskip
\noindent\textit{Data availability}---No data were created or analyzed in this study.

\bibliographystyle{apsrev4-2}
\bibliography{prl_submission_refs}

\end{document}